\documentclass[aps,prl,twocolumn,nofootinbib]{revtex4}

\usepackage{graphicx}
\usepackage{amssymb, amsmath,mathtools}
\usepackage{bbm}
\usepackage{enumerate}
\usepackage{bbold}

\graphicspath{{./}{./plots/}}

\usepackage{xcolor}

\begin{document}

\title{Bogoliubov-Fermi Surfaces in Noncentrosymmetric Multicomponent Superconductors}

\author{Julia M. Link and Igor F. Herbut}

\affiliation{Department of Physics, Simon Fraser University, Burnaby, British Columbia, Canada V5A 1S6}

\begin{abstract}

We show that when the time reversal symmetry is broken in a multicomponent superconducting condensate without inversion symmetry
the resulting Bogoliubov quasiparticles generically exhibit mini-Bogoliubov-Fermi (BF) surfaces, for small superconducting order parameter.
The absence of inversion symmetry makes the BF surfaces stable with respect to weak perturbations.
With sufficient increase of the order parameter, however, the Bogoliubov-Fermi surface may disappear through a Lifshitz transition, and the spectrum this way become fully gapped. Our demonstration is based on the computation of the effective Hamiltonian for the bands near the normal Fermi surface by the integration over high-energy states. Exceptions to the rule, and experimental consequences are briefly discussed.

\end{abstract}

\maketitle

The appearance of the gap in the quasiparticle spectrum of an $s$-wave superconductor has been one of the defining features of the superconducting state of matter  since the conception of the theory of Bardeen, Cooper, and Schrieffer \cite{schrieffer}. Many unconventional superconductors of today do not feature a full gap, but still reduce the density of quasiparticle states near the Fermi energy by leaving only lines or points in the momentum space where the gap vanishes \cite{volovik}. These, however, are not the only possibilities \cite{yang, Wilczek, Gubankova}, and Fermi surfaces of Bogoliubov excitations in the superconducting state are possible as well \cite{agterberg, brydon, Setty2020}. These arise in superconductors with more than one band participating in pairing, and when the condensate breaks time reversal (TR) symmetry while preserving inversion, which is present in both normal and superconducting phases. The presence of inversion symmetry has been deemed crucial for the appearance and particularly the stability of a Boguliubov-Fermi (BF) surface, which then comes out topologically protected. The existence of a surface of gapless quasiparticle excitations leads to a finite residual density of states, and has many consequences for the low temperature properties of the superconducting phase. It should be detectable in the temperature dependence of the penetration depth, heat conductivity, and heat capacity at low temperatures, for example \cite{timm}.

It has been recently found in an example of TR-symmetry-breaking superconducting ground state in a topologically nontrivial (Rarita-Schwinger-Weyl) four-band system that the BF surfaces can form despite the complete lack of inversion symmetry in the superconducting states \cite{link}. Other instances of the same phenomenon have also been considered \cite{timm1, sim}. The generality of the emergence of the BF surfaces in materials with no inversion symmetry  has not been clear, however, and its possible relevance to the large number of known noncentrosymmetric superconductors \cite{smidman} is an open issue.

In this Letter we show that in a multiband system without inversion symmetry spontaneous breaking of TR in the superconducting state generically leads to the formation of a BF surface, at least right below the critical temperature, if the superconducting phase transition is continuous. With an increase of the order parameter the BF surface may eventually shrink to a point and then be replaced by a gap. The latter phenomenon occurs when the superconducting gap becomes comparable to the energy gap between the band at and the bands away from the normal Fermi surface. Central to our demonstration is the derivation of the effective low-energy Hamiltonian, which may be thought of as a result of integrating out the energy bands away from the Fermi energy. It provides more than just a useful approximate picture of the spectrum of Bogoliubov quasiparticles, as we show that the location of the zero modes of the effective Hamiltonian in the momentum space coincides with the location of the BF surface of the original Bogoliubov-de Gennes (BdG) quasiparticle Hamiltonian of arbitrary size. The absence (or presence) of the TR symmetry in the superconducting state governs the algebra behind the computation of the effective Hamiltonian, and dictates the low-energy spectrum. The breaking of the TR symmetry leads to mini BF surfaces, while the preservation of the TR symmetry leads generically to gapless lines of Bogoliubov quasiparticles, similarly as in inversion-symmetric systems \cite{sigrist, boettcher}.

{\it Boguliubov-de Gennes Hamiltonian.}--Consider the quantum-mechanical action for the Bogoliubov quasiparticles in the superconducting state:
\begin{equation}
S= k_B T \sum_{\omega_n, \textbf{p}} \Psi^\dagger (\omega_n, \textbf{p}) [-i\omega_n + H_{\rm{BdG}} (\textbf{p}) ]  \Psi(\omega_n, \textbf{p}),
\label{eq:action_Eq1}
\end{equation}
where the Nambu spinor is ${ \Psi (\omega_n, \textbf{p}) =  \big[\psi(\omega_n,\textbf{p}) , \mathcal{T} \psi(\omega_n, \textbf{p} ) \big] ^{\rm T} }$, $\textbf{p}$ is the momentum, $\omega_n = (2n+1) \pi k_B T$ is the Matsubara frequency, and $T$ is the temperature. $\psi=(\psi_1,\cdots,\psi_N)$ is a $N$-component Grassmann number describing $N$ energy bands, and its time reversed counterpart is $\mathcal{T} \psi(\omega_n, \textbf{p} )= U \psi^* (-\omega_n, -\textbf{p})$, where $\mathcal{T}$ is the antiunitary time-reversal operator, and $U$ its unitary part. This way the BdG Hamiltonian becomes:
\begin{eqnarray}
 H_{\rm BdG}(\textbf{p}) &=&
 \begin{pmatrix} H(\textbf{p})-\mu   & \Gamma  \\
 \Gamma^\dagger & - \big[ H(\textbf{p})- \mu  \big]
 \end{pmatrix}
 \:.
 \label{eq:BdG-Ham-Eq2}
\end{eqnarray}
We assume that the $N$-dimensional Hamiltonian $H(\textbf{p})$ is only TR symmetric, so that $U^\dagger H(\textbf{p}) U= H^* (-\textbf{p})$. Recalling that it is also Hermitian and $ H^* (\textbf{p})= H^{\rm T} (\textbf{p})$, the action in Eq.~(\ref{eq:action_Eq1})  would assume its textbook form.

For simplicity we also assume that the $N$-dimensional matrix $\Gamma$ which denotes the intra- and interband Cooper pairing is constant, so that the pairing term is local in real space, $\sim \Psi^\dagger (\textbf{x}, \tau) \Gamma  (\mathcal{T} \Psi (\textbf{x},\tau)) $. The matrix $\Gamma$ can then be expanded as $\Gamma= \sum_a \Delta_a M_a$, with index $a$ labeling the complex order parameter components, $ \Delta_a= \Delta_{1a}+ i \Delta_{2a}$. $M_a$ are Hermitian matrices that form a basis in the relevant order parameter space, and need only to conform to the fermionic statistics of the fields.  If $(\mathcal{T})^2 = -1$, fermionic statistics dictates that all $M_a$ are even under TR, so $s$-wave and tensorial $d$-wave order parameters are among the obvious examples\cite{boettcher, igorprd}. In the case of pairing of the (effective) integer-spin fermions for which $(\mathcal{T})^2 =+1$, the matrices $M_a$ would be TR odd, like three components of a  $p$-wave\cite{sim}. Our method will work for both cases, and can also be easily generalized to momentum-dependent pairing.

The BdG Hamiltonian in Eq.~\eqref{eq:BdG-Ham-Eq2}  can thus also be written as
\begin{equation}
H_{\rm BdG} = \sigma_3 \otimes [H(\textbf{p}) -\mu] + \sum_a( \Delta_{1a} \sigma_1 \otimes M_a - \Delta_{2a} \sigma_2 \otimes M_a)
 \:,
\end{equation}
where $\sigma_\alpha$, $\alpha=1,2,3$ are the usual Pauli matrices.
The phase common to all $\Delta_a$ is assumed to have been gauged away.
If $M_a$ is TR even, $H_{\rm BdG}$ is even under the time-reversal operator
$\mathbb{1}_{2 \times 2}\otimes \mathcal{T}$ only when all $ \Delta_{2a}=0$. If some $\Delta_{2a} \neq 0$, and, consequently
$\Gamma \neq \Gamma^ \dagger$, TR is broken in the superconducting phase. For completeness, let us also consider the case when
 $\mathcal{T}^2 = +1$ when $M_a$ are odd: $H_{\rm BdG}$ will then be even under
$\mathbb{1}_{2\times 2}\otimes \mathcal{T}$ when all $\Delta_{1a}  = 0$. One can then still gauge away the overall phase of $\pi/2$ to have the pairing matrix $\Gamma$ Hermitian. For either type of the TR symmetry, non-Hermiticity of the pairing matrix $\Gamma$ is thus tantamount to breaking of the TR symmetry in the superconducting state.

{\it Effective Hamiltonian.}--Let us define the eigenvalues (bands) and the eigenstates of the normal state Hamiltonian $H(\textbf{p})$,  as $E_i(\textbf{p})$ and $ \phi_i (\textbf{p})$, $i=1,...N$. We consider a momentum $\textbf{p}$ at the normal state's Fermi surface at which there is only one eigenvalue equal to the chemical potential $\mu$, since the normal Fermi surface is in general nondegenerate, except possibly at special points. There could be more than one connected Fermi surface, but without inversion there is no forced double degeneracy of the Fermi surface at all momenta $\textbf{p}$; the TR alone implies only that if a momentum $\textbf{p}$ belongs to the Fermi surface, the opposite momentum  $-\textbf{p}$ does as well. We may call the eigenstate with its energy arbitrarily close to the Fermi surface $\phi_1 (\textbf{p})$ ``light", and the remaining $N-1$ eigenstates ``heavy". This separation may depend on the Fermi surface point under consideration.

The spectrum of the Bogoliubov quasiparticles at a momentum $\textbf{p}$ is given by the solution of the equation for the real frequency $\omega$
\begin{equation}
\det [H_{\rm BdG}(\textbf{p})-\omega ] =0
\:.
\label{eq:Especrum-BdG_Eq4}
\end{equation}
With the separation into light and heavy states at a given momentum near the normal Fermi surface, one can write the BdG Hamiltonian in the
basis $\{  (\phi_i (\textbf{p}),0)^T, (0,\phi_i (\textbf{p}))^T \}$, $i=1,...N$ as
\begin{eqnarray}
 H_{\rm BdG}(\textbf{p}) &=&
 \begin{pmatrix} H_l (\textbf{p}) & H_{lh}(\textbf{p})  \\
 H_{lh}^\dagger (\textbf{p}) & H_h (\textbf{p})
 \end{pmatrix}
 \:.
 \label{eq:Ham_light-heavy-modes_Eq5}
\end{eqnarray}
 The block for the light particle and hole states $H_l(\textbf{p})$ is a two-dimensional matrix. The heavy modes are described by the $(2N-2)$-dimensional matrix $H_h (\textbf{p}) $, and the coupling between the light and heavy states $H_{lh}(\textbf{p})$ is a $2\times (2N-2)$ matrix. The above determinant can now be rewritten as
 \begin{equation}
\det [H_{\rm BdG}(\textbf{p}) -\omega ]= \det [H_{h}(\textbf{p}) -\omega ] \det L_{ef}(\omega, \textbf{p})
\:,
\label{eq:Schurcomplement-Eq6}
\end{equation}
 where the effective Lagrangian  $L_{ef}$ is the {\it Schur complement} \cite{schur} of the block matrix for the heavy modes:
 \begin{equation}
 L_{ef} (\omega,  \textbf{p}) = H_l (\textbf{p}) -\omega - H_{lh}(\textbf{p}) (H_h (\textbf{p}) -\omega )^{-1} H_{lh}^\dagger (\textbf{p}).
 \label{eq:Schurcomplement-Eq7}
\end{equation}
 The first factor in Eq.~\eqref{eq:Schurcomplement-Eq6} may be understood as the fermionic partition function at a fixed frequency for the heavy modes, and the second factor is therefore the residual partition function (at fixed frequency) for the light modes, which are modified by the integration over the heavy modes \cite{supplement}. $L_{ef} (\omega, \textbf{p})$ is defined whenever the heavy block is invertible, which is the case if
 $|\omega|< |E_i (\textbf{p})-\mu| $ for $i \neq 1$. Under this condition the eigenvalue equation in Eq.~\eqref{eq:Especrum-BdG_Eq4}  reduces to
$
\det  L_{ef}(\omega, \textbf{p}) =0.
$
Although $L_{ef}(\omega, \textbf{p})$ is only a two-dimensional matrix, its computation involves an inversion of the $(2N-2)$-dimensional matrix, so for a general $\omega$ there is no obvious gain. $\omega=0$, however, is a solution only when
\begin{equation}
\det  H_{ef}(\textbf{p}) =0,
\label{eq:determinant_eff_Ham_Eq9}
\end{equation}
with $H_{ef} (\textbf{p}) = L_{ef} ( 0,   \textbf{p}  )$, and which may be called the effective Hamiltonian \cite{berg}. We emphasize that only the solutions for zero modes of $H_{ef}(\textbf{p})$ are exactly the same as those for the original $H_{\rm BdG}(\textbf{p})$; the rest of their spectra differs. This is, however, all that is needed to understand the emergence of the BF surface, as we show next.

\begin{figure*}%[t]
\centering
\includegraphics[width=2\columnwidth]{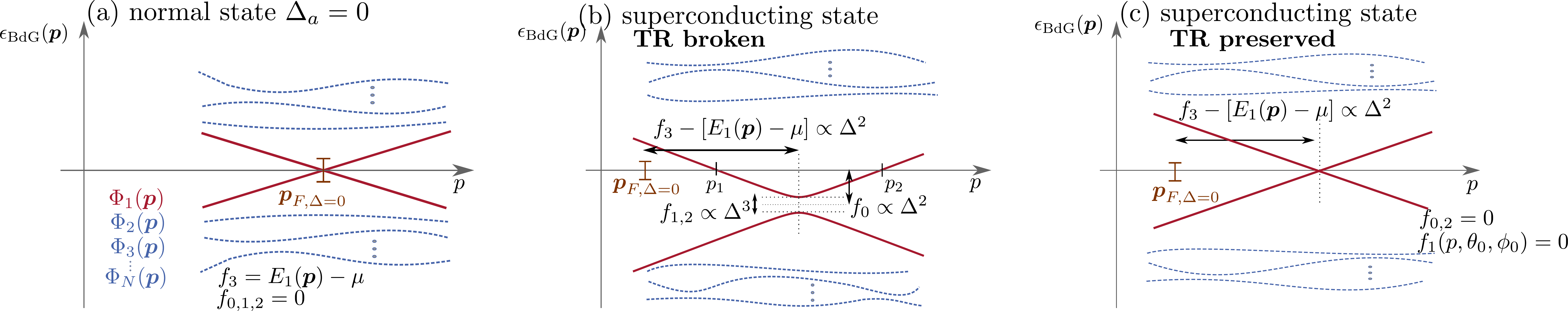}
\caption{(a) The energy dispersion of the Bogoliubov quasiparticles in the direction orthogonal to the Fermi surface of the light (red) and the heavy (blue) particle and hole states. (b) The same in the superconducting state with broken TR symmetry, in the direction $(\theta_0,\phi_0)$ where the first order contribution to the gap vanishes. The energy dispersion of the light states is shifted in momentum and energy by the amount $\mathcal{O}(\Delta^2)$, the light particle and hole states are mixed by the term of the order $\mathcal{O}(\Delta^3)$, and as a result the energy of the Bogoliubov quasiparticles vanishes at some $p_1$ and $p_2$. Varying the direction in the momentum space away from $(\theta_0,\phi_0)$ smoothly changes the solutions $p_1$ and $p_2$, until they merge and that way close a surface. (c) The energy dispersion of the Bogoliubov quasiparticles in the superconducting state with preserved TR symmetry, in the direction $(\theta_0,\phi_0)$. The energy dispersion of the light states is only shifted in the momentum direction, which leads to line nodes.}
\label{Fig1}
\end{figure*}

{\it Bogoliubov-Fermi surface.}--The effective Hamiltonian is two dimensional and thus may be expanded in the Pauli basis
\begin{equation}
H_{ef} = \sum_{\alpha=0} ^3 f_{\alpha} ( \textbf{p} ) \sigma_{\alpha}
\end{equation}
with $\sigma_0 = \mathbb{1}_{2 \times 2}$. Equation~\eqref{eq:determinant_eff_Ham_Eq9} can now be written as
\begin{equation}
f_0  ^2 (\textbf{p} )-\sum_{i=1}^3 f_{i} ^2(\textbf{p} ) = 0.
 \end{equation}
We will show that if in some direction in the momentum space the last equation is solved by {\it two} different magnitudes $p=p_1$ and $p=p_2$, the emergence of a BF surface follows from continuity: varying the direction smoothly changes the solutions $p_1$ and $p_2$, until they merge and that way close a surface.

 In the normal phase, when $\Delta_a \equiv 0$ all the states are decoupled, and there is of course the normal Fermi surface at which  $f_3  ( \textbf{p} )=E_1 (\textbf{p}) - \mu $ changes sign, and  $f_{\beta} ( \textbf{p} )\equiv 0$, for $\beta=0,1,2$ [see Fig.\ref{Fig1}(a)].

 We will find that in the TR-symmetry-breaking superconducting phase $f_0 = \mathcal{O}(\Delta^2)$, $f_3  ( \textbf{p} )-[E_1 (\textbf{p}) - \mu]= \mathcal{O}(\Delta^2) $, whereas $f_{1,2} = \mathcal{O}(\Delta) + \mathcal{O}(\Delta^3) $. $\Delta^2 = \sum_a |\Delta_a|^2 $
 is an overall size of the order parameter. A finite value of $f_0$ introduces a shift of the order $\mathcal{O}(\Delta^2)$ in the energy of the light particle and hole states due to the interband pairing, $f_3  ( \textbf{p} )-[E_1 (\textbf{p}) - \mu]$ introduces a shift in the momentum direction also of the order $\mathcal{O}(\Delta^2)$, and $f_{1,2}$ open a gap between the light particle and light hole state of the order $\mathcal{O}(\Delta) + \mathcal{O}(\Delta^3)$.
 Whenever the leading $\mathcal{O}(\Delta)$ contributions to $f_{1,2}$ vanish somewhere on the normal Fermi surface, i.e., the intraband coupling between the light particle and hole state vanishes, there will be two different points at $p_1$ and $p_2$ where the energy of the quasiparticles is equal to the chemical potential and the BF surface will be nucleated in the superconducting phase provided $\Delta$ is small enough. The vanishing of the leading order contribution to $f_{1,2}$ yields two conditions on two polar angles, so the BF surfaces in form of an inflated point node will in general emerge around particular points near the normal Fermi surface. If the two conditions on the polar angles happen to be the same, the form of the BF surface will be an inflated line node. The principle behind the emergence of the BF surface is depicted in Fig.~\ref{Fig1}.

 When the TR is preserved in the superconducting state, on the other hand, $f_i (\textbf{p} ) \equiv 0$ for $i=0,2$; there is no shift in the energy of the light particle and hole states, but only a shift in the momentum. This implies $p_1 = p_2$. Zero-energy solutions are then given by $f_i (\textbf{p})=0$ for $i=1,3$, which provides two conditions on three variables, and leads to a line of gapless points \cite{schnyder}.

{\it Iterative procedure.}--To see how this comes about let us write the BdG Hamiltonian in Eq.~\eqref{eq:Ham_light-heavy-modes_Eq5} once again as
\begin{eqnarray}
H_{{\rm BdG},N} ^{(0)}  &=&
\begin{pmatrix} H_{1,1}^{(0)} & H_{1,2}^{(0)} & \ldots & H_{1,N}^{(0)}  \\
 H_{1,2}^{(0) \dagger} & H_{2,2}^{(0)} & \ldots  &  H_{2,N}^{(0)} \\
 \vdots & \vdots & \ddots & \vdots  \\
 H_ {1,N}^{(0) \dagger} & H_{2,N}^{(0) \dagger} & \ldots & H_{N,N}^{(0)}
 \end{pmatrix}
 \:,
 \label{eq:BdGham_light_heavy_Eq11}
\end{eqnarray}
with the two-dimensional blocks as
\begin{equation}
H_{k,m}^{(0)}= \delta_{k,m} [E_k( \textbf{p})-\mu ] \sigma_3  + \sum_{a} (\phi^\dagger _k M_a \phi_m) ( \Delta_{1a} \sigma_1 - \Delta_{2a} \sigma_2).
\end{equation}
Note that the diagonal blocks are Hermitian matrices whereas the off-diagonal blocks in general are not.

Using the Schur decomposition \cite{schur} again,
\begin{equation}
\det H_{{\rm BdG},N}^{(0)} = \det H_{N,N}^{(0)} \det H_{{ \rm BdG}, N-1} ^{(1)},
\end{equation}
where $H_{{\rm BdG}, N-1}^{(1)}$ is the Schur complement of the last block on the diagonal $H_{N,N} ^{(0)}$,
%
%\begin{widetext}
\begin{eqnarray}
 \label{eq:Schur-complement_Eq15}
 H_{{\rm BdG}, N-1}^{(0)}- H_{{\rm BdG},N-1} ^{(1)}=\quad& &\\
 \begin{pmatrix} H_{1,N}^{(0)}\\
 H_{2,N}^{(0)}\\
 \vdots \\
 H_ {N-1,N}^{(0)}
 \end{pmatrix}
 \cdot
 \big(H_{N,N}^{(0)}\big) ^{-1}
 \cdot
 \big(H_{1,N} ^{(0) \dagger}, H_{2,N}^{(0) \dagger},\ldots,H_{N-1,N}^{(0) \dagger} \big)
 \:, \nonumber
\end{eqnarray}
%\end{widetext}
%
and as a matrix it consists of $(N-1)\times (N-1) $ two-dimensional blocks. One can think of it as the effective Hamiltonian for the $N-1$ bands after only the $N$th band has been integrated out.  This step can be now iterated so that
\begin{widetext}
\begin{eqnarray}
\det H_{{\rm BdG},N}^{(0)} =
\det H_{N,N}^{(0)} \det H_{N-1,N-1} ^{(1)} \det H_{N-2,N-2} ^{(2)}...\det H_{1,1} ^{(N-1)},
\end{eqnarray}
\end{widetext}
where each matrix $H_{N-k,N-k} ^{(k)}$ is a two-dimensional ``heavy" diagonal block of the effective BdG Hamiltonian at the (intermediate) $k$th stage of  the iteration, and the requisite effective Hamiltonian $H_{ef} $ for the light states is simply the final $H_{1,1} ^{(N-1)}$. This way no inversion of anything larger than a two-dimensional matrix is ever required, but the price is the tracking of the evolution of the parameters appearing in the effective Hamiltonians of the reduced size.

{\it Results.}--What is the result of this procedure? If the TR is preserved and $M_a$ is TR even one can set  $\Delta_{2a}\equiv 0$. Equation~\eqref{eq:Schur-complement_Eq15} implies that at each iteration one multiplies three matrices that are linear combinations of only $\sigma_1$ and $\sigma_3$. Such a multiplication can yield only another linear combination of the same $\sigma_1$ and $\sigma_3$, since
$
\rm{Tr} (\sigma_\mu \sigma_i \sigma_j \sigma_k )\equiv 0
$
if $i,j,k = 1,3$ and $\mu=0,2$.  All the blocks $H_{N-k,N-k} ^{(k)}$  are thus real and traceless, and therefore in the final effective Hamiltonian $f_\alpha (\textbf{p}) \equiv 0$ for $\alpha=0,2$ at every momentum $\textbf{p}$ as well. The solution of two equations $f_\beta (\textbf{p})=0$, for $\beta=1,3$ will then, in general, lead to lines of gapless points in the momentum space.

When TR is broken in the superconductor, already after the first iteration all the blocks in the $H_{{\rm BdG}, N-1}^{(1)}$ become unrestricted general $2\times 2$ matrices, and remain so at further iterations. So $f_\alpha (\textbf{p})\neq 0$ for $\alpha=0,1,2,3$. The BF surface may result, however, when the superconducting order is sufficiently weak. Equation~\eqref{eq:Schurcomplement-Eq7} implies that the correction to $H_l$  in the $H_{ef}$ is quadratic in $H_{lh}$. Since $H_{lh}\sim \Delta $, the leading order correction is of second order in the superconducting order parameters. To this order one may therefore neglect all off-diagonal, $\sim\Delta$, matrix elements in $H_h$ in Eq.~\eqref{eq:Schurcomplement-Eq7}, and in this way arrive at the familiar expression from the second-order perturbation theory. In the notation of Eq.~\eqref{eq:BdGham_light_heavy_Eq11} then
\begin{equation}
H_{ef} = H_{1,1}^{(0)} - \sum_{k=2}^N H_{1,k}^{(0)} \big(H_{k,k;\Delta_a=0}^{(0)}\big) ^{-1}  H_{1,k}^{(0) \dagger} + \mathcal{O}(\Delta^3).
\end{equation}
The crucial observation is that since $H_{1,k}^{(0)}$ for $k\neq 1$ are off-diagonal [i. e., linear combinations of only $\sigma_1$ and $\sigma_2$, as in Eq. (12)], and $H_{k,k;\Delta_a=0}^{(0)}$ (and therefore its inverse) are proportional to $\sigma_3$, the leading order correction in $H_{ef}$ is diagonal, i. e., a linear combination only of $\sigma_3$ and unit matrix. The functions $f_\mu$, $\mu=1,2$ therefore do not acquire an $\mathcal{O}(\Delta^2)$ correction, and
\begin{equation}
f_1 (\textbf{p}) - i f_2 (\textbf{p}) = \phi_1 ^\dagger (\textbf{p}) \Gamma  \phi_1 (\textbf{p}) + \mathcal{O}(\Delta^3).
\end{equation}
In contrast, $f_{\mu}$ for $\mu=0,3$, do. Explicitly, \cite{supplement}
\begin{equation}
f_3 (\textbf{p})= E_1 (\textbf{p}) - \mu + \sum_{k=2}^N \frac{ |\phi_1 ^\dagger (\textbf{p}) \Gamma \phi_k (\textbf{p})|^2 + |\phi_1 ^\dagger (\textbf{p})\Gamma^\dagger \phi_k (\textbf{p}) |^2  } {2 [E_k(\textbf{p})-\mu]}
\end{equation}
and, most importantly,
\begin{equation}
f_0 (\textbf{p}) =  \sum_{k=2}^N \frac{ |\phi_1 ^\dagger (\textbf{p}) \Gamma \phi_k (\textbf{p}) |^2 - |\phi_1 ^\dagger (\textbf{p}) \Gamma^\dagger \phi_k (\textbf{p}) |^2  } {2 [E_k(\textbf{p}) - \mu] }
\:,
\end{equation}
with the next-order terms in the last two equations being $\mathcal{O}(\Delta^4)$.

At the points on the normal Fermi surface where
\begin{equation}
\phi_1 ^\dagger (\textbf{p}) \Gamma  \phi_1 (\textbf{p})=0,
\end{equation}
the off-diagonal elements $f_{1,2}$ of $H_{ef}$ become $\mathcal{O}(\Delta^3)$ and negligible, so the leading effect of heavy modes is to shift the energy bands in the momentum and energy directions, as in Fig.~\ref{Fig1}. This inevitably leads to two momenta near the original Fermi momentum at which $H_{ef}$ has zero-energy eigenstates. The BF surface is then nucleated around that particular point on the normal Fermi surface by continuity.

The last equation may not have a solution, in which case the spectrum will be gapped. One such instance is when $\Gamma = \Delta_1 M_1 + i \Delta_2 M_2$, with $M_1 = \mathbb{1}_{N \times N}$, i. e., the real part is the $s$-wave. If neither $f_1 (\textbf{p})$ nor $f_2 (\textbf{p})$ are simple constants, however, the equation will typically have several solutions, and the BF surfaces will ensue. An example is provided by the quasiparticle spectrum of some of the $d$-wave superconducting states in  the Rarita-Schwinger-Weyl semimetals \cite{link}.

{\sl Discussion.}--The BF surface once nucleated is fully stable to weak perturbations. This is because no symmetry is left in the superconducting state, apart from the translational symmetry, that could be broken. This is an important difference from the standard case with inversion, where inversion symmetry is susceptible to spontaneous breaking by favorable residual interactions in the superconducting state \cite{oh}. The final result of such interaction-induced reduction of inversion symmetry would be precisely the stable BF surface discussed here.

Increasing sufficiently the superconducting order parameter would shrink the BF surface to a point, and replace it by a gap. Such a transition is not accompanied by breaking of any symmetry, however, and provides an example of a Lifshitz transition \cite{lifshitz}. It occurs at $\Delta_c ^2 = (E_2 (\textbf{p}) -\mu)^2 /B(\textbf{p}) $, if $|E_2 (\textbf{p}) -\mu| \ll |E_k (\textbf{p}) -\mu|$, for $k>2$, for example, with the number $B (\textbf{p}) \sim 1$ \cite{supplement}. Such a characteristic energy scale $|E_2 (\textbf{p}) -\mu|$ in noncentrosymmetric superconductors usually originates from the asymmetric spin-orbit coupling (ASOC), $|E_2 (\textbf{p}) -\mu|\sim E_{\rm ASOC}$ and typically $(\Delta/ E_{\rm ASOC} )^2 \ll 1$  \cite{smidman}. Very small such a ratio, on the other hand, is detrimental for the size of the BF surface, which is $\sim \Delta^2/ |E_2 (\textbf{p}) -\mu|$, for $\Delta/ |E_2 (\textbf{p}) -\mu| \ll 1$.  A crude estimate gives the largest BF surface for the ratio $\Delta \approx \Delta_c /\sqrt{2}$ \cite{supplement}.

The systems with inversion in both normal and superconducting states \cite{agterberg} may be studied in analogy with the present calculation. The effective Hamiltonian is then four-dimensional, however, which introduces further subtleties in the algebra behind its computation. One may, nevertheless, understand the appearance and the stability of the BF surface in that case without resorting to topology. The details of this approach will be presented in a separate publication.

Examples of noncentrosymmetric superconductors with broken TR are believed to include LaNiC${}_2$ \cite{smidman,Hillier,Lee,Bondale,Pecharsky,Chen,Iwamoto}, LaNiGa${}_2$ \cite{weng}, La${}_7$Ir${}_3$ \cite{Barker}, and Re${}_6$Zr \cite{Singh}. All four materials are also commonly assumed to be fully gapped, however. It would be interesting to identify a noncentrosymmetric material that breaks TR but displays $\sim T^3 $ behavior in the specific heat over a range of temperatures, for example. Instead of extending all the way to zero we would predict this behavior crossing over to $\sim T$  below $\sim T_c ^2 /E_{\rm ASOC}$, provided that the BF surface survives. Similar crossovers that would reflect a finite residual density of states in the superconducting phase should be observable in the penetration depth and thermal conductivity as well.

\begin{acknowledgments}
We are grateful to I. Boettcher and C. Timm for many useful discussions and comments. J.M.L. is supported by the  DFG Grant No. LI 3628/1-1, and I.F.H. by the NSERC of Canada.
\end{acknowledgments}
\newpage

\setcounter{equation}{0}
\setcounter{page}{1}
\begin{widetext}
 \section{Supplementary material}
We present an alternative derivation of the Schur decomposition of the determinant of the Bogoliubov - de Gennes Hamiltonian. We demonstrate that this decomposition (Eq.~(6) in the paper) can also be arrived at by the (Gaussian) integration over heavy modes in the partition function for fixed frequency. In addition, an explicit derivation of the effective Hamiltonian in a simple example which illustrates and further elaborates the main results of the paper is provided.
\section{Mode elimination}
An alternative to the standard way \cite{schur} of arriving at Eqs.~(6) and ~(7) is the (Gaussian) integration over the heavy modes  in the partition function for a fixed frequency. Let us write such a partition function defined by the Hamiltonian in Eq.~(5):
\begin{eqnarray}
 \mathcal{Z}(\omega)&=&\int_{\psi_l, \psi_h}
 e^{-\big( \psi_l^\dagger ( H_l  -\omega ) \psi_l + \psi_h^\dagger (H_h -\omega )\psi_h  +\psi_h^\dagger H_{lh}^\dagger \psi_l + \psi_l^\dagger H_{lh} \psi_h \big)}.
 \end{eqnarray}
 The integration variables could be complex or Grassmann, and the outcome would be the same. We choose Grassmann here since it is closer to the physics of the problem.  The partition function can be rearranged as
 \begin{eqnarray}
 \mathcal{Z}(\omega)  = \int_{\psi_l} e^{- \psi_l^\dagger [H_l -\omega - H_{lh} (H_h -\omega ) ^{-1} H_{lh} ^\dagger ]  \psi_l}
 \times
 \int_{\psi_h} e^{-  \big(\big[\psi_h^\dagger + \psi_l ^\dagger H_{lh} (H_h -\omega )^{-1} \big]  [H_h-\omega]   \big[\psi_h +   (H_h -\omega )^{-1} H_{lh} ^\dagger \psi_l \big] \big) }
 \:.
\end{eqnarray}
Changing the Grassmann integration variables in the second integral  as
\begin{equation}
\tilde{\psi}_h = \psi_h +   (H_h -\omega )^{-1} H_{lh} ^\dagger \psi_l,
\end{equation}
we thus readily find
\begin{equation}
 \mathcal{Z}(\omega) = \mathcal{Z}_{\rm eff}  (\omega) \int_{\tilde{\psi}_h} e^{- \tilde{\psi}_h^\dagger(H_h -\omega ) \tilde{\psi}_h},
\end{equation}
with the effective partition function of the light fermions
\begin{equation}
 \mathcal{Z}_{\rm eff}(\omega) = \int_{\psi_l}  e^{- \psi_l^\dagger L_{ef}(\omega,\textbf{p})\psi_l }.
\end{equation}
and with the $L_{ef} (\omega,\textbf{p}) $ as defined by Eq.~(7). The result in Eq.~(6) follows.

\section{Two-band system}
Let us a assume a system with just two energy bands ($N=2$), without inversion symmetry, defined by the normal state Hamiltonian $H (\textbf{p}) $. Without inversion, at a generic value of the momentum there exists only one eigenstate $\phi_i (\textbf{p})$ at a given energy, in contrast to the systems with inversion. Let us assume that the energy of the eigenstate $\phi_1 (\textbf{p}) $ of the normal state Hamiltonian is  at the given momentum equal to the chemical potential $\mu$, and call this state the light state. The remaining $\phi_2 (\textbf{p})$ is then the heavy state. This labeling may depend on the momentum, as illustrated in Fig.~\ref{fig:SM_fig1}.
\begin{figure}
\includegraphics[width=0.5\columnwidth]{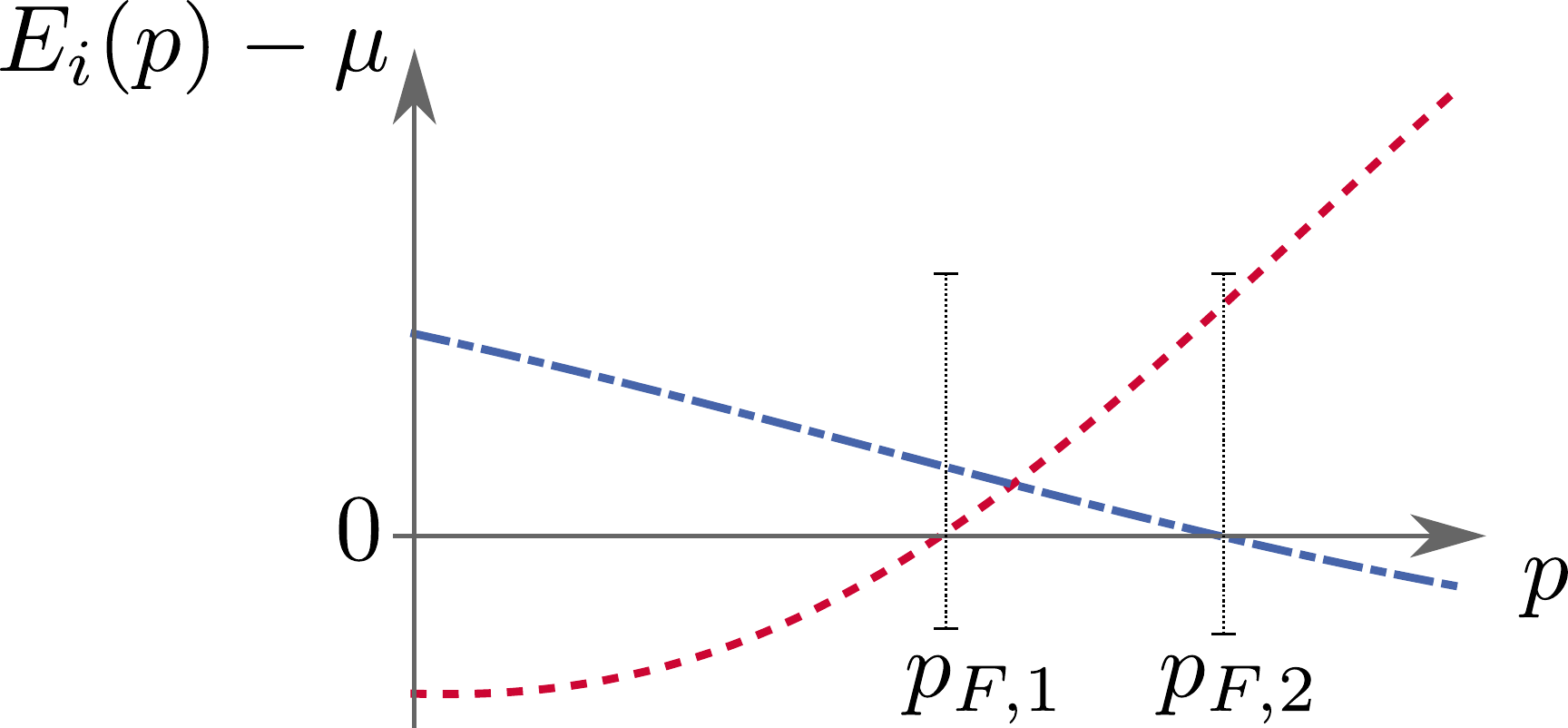}
\caption{An example of two energy bands of the normal state Hamiltonian $H(\textbf{p})$ dependent on the momentum $p$. There are two normal Fermi surfaces at $p_{F,1}$ and $p_{F,2}$. To study the emergence of the BF surface at $p_{F,1}$, the red, dashed energy band is the light band $E_1(\textbf{p})$ with the eigenstate $\phi_1 (\textbf{p})$ and the blue, dash-dotted line is the heavy band $E_2(\textbf{p})$ with $\phi_2(\textbf{p})$. However, if the emergence of the BF surface at $p_{F,2}$ is being considered, the roles of the bands are exchanged and the blue, dashed-dotted curve is the light energy band with $E_1(\textbf{p})$ and $\phi_1(\textbf{p})$, while the red, dashed curve is the heavy state $E_2(\textbf{p})$ with $\phi_2(\textbf{p})$. }
\label{fig:SM_fig1}
\end{figure}

This minimal Hamiltonian may be understood as the dominant block when other heavy states are much heavier than the state ``2", i. e. $|E_2 (\textbf{p}) -\mu| \ll |E_k (\textbf{p}) -\mu|$, for $k=3,...$. This is physically the case when the energy split  $|E_2 (\textbf{p}) -E_1 (\textbf{p})|$ derives from the asymmetric spin-orbit coupling, for example. Furthermore, we will see shortly that the heavier states (with $k>2$), when present in the original normal state Hamiltonian,  after the elimination process can only provide $\mathcal{O}(\Delta^2)$ corrections to the diagonal, and $\mathcal{O}(\Delta^3)$ to the off-diagonal terms in all four $2\times 2$ blocks in our starting BdG Hamiltonian given below. To the leading order in the effective Hamiltonian the contributions of different heavy states are simply additive. It therefore suffices to consider a single heavy state to illustrate our main points.

The Bogoliubov - de Gennes (BdG) Hamiltonian for $N=2$ can be written as

\begin{equation}
 H_{{\rm BdG},2}(\textbf{p})=
 \begin{pmatrix}
  H_{l}(\textbf{p}) & H_{lh}(\textbf{p})\\
  H_{lh}(\textbf{p})^\dagger & H_h(\textbf{p})
 \end{pmatrix}
=
\begin{pmatrix}
 H_{1,1}^{(0)} & H_{1,2}^{(0)}\\
 H_{1,2}^{(0)\dagger} & H_{2,2}^{(0)}
\end{pmatrix}
\label{eq:BdGexample}
 \end{equation}
with
\begin{eqnarray}
 H_l(\textbf{p})&=&
 H_{1,1}^{(0)}
 =
 \begin{pmatrix}
  E_1(\textbf{p}) - \mu & x (\textbf{p})\\
  \bar{x}(\textbf{p}) & -[E_1(\textbf{p}) - \mu ]
 \end{pmatrix}
\quad \text{ with }
x(\textbf{p})=\phi_{1}^\dagger(\textbf{p}) \Gamma \phi_{1}(\textbf{p}) \quad \text{ and } \bar{x}(\textbf{p})=\phi^\dagger_1(\textbf{p}) \Gamma^\dagger \phi_1(\textbf{p})
\:,\\
 H_h(\textbf{p})&=&
 H_{2,2}^{(0)}
 =
 \begin{pmatrix}
  E_2(\textbf{p}) - \mu & y(\textbf{p})\\
  \bar{y}(\textbf{p}) & -[E_2(\textbf{p}) - \mu ]
 \end{pmatrix}
\quad \text{ with }
y(\textbf{p})=\phi_{2}^\dagger (\textbf{p}) \Gamma \phi_{2}(\textbf{p})
\:,\\
 H_{lh}(\textbf{p})&=&
 H_{1,2}^{(0)}
 =
 \begin{pmatrix}
  0 & u(\textbf{p})\\
  v(\textbf{p}) & 0
 \end{pmatrix}
\quad \text{ with }
u(\textbf{p})=\phi_{1}^\dagger (\textbf{p}) \Gamma \phi_{2}(\textbf{p}) \quad \text{ and } \quad
v(\textbf{p})=\phi_{1}^\dagger (\textbf{p}) \Gamma^\dagger \phi_{2}(\textbf{p})
\:.
\end{eqnarray}
The matrix-blocks $H_{i,j}^{(0)}$ in the Hamiltonian Eq.~\eqref{eq:BdGexample} are two-dimensional. This is the crucial difference with the inversion-symmetric case: inversion symmetry implies that the  Fermi surface is doubly degenerate at all momenta, which leads to 4-dimensional matrices $H_{i,j}^{(0)}$.
%%%

 After we integrate out the heavy state ``2", the effective Hamiltonian becomes
\begin{eqnarray}
 H_{ef} &=&
 H_{1,1}^{(0)}-H_{1,2}^{(0)} (H_{2,2}^{(0)})^{-1} H_{1,2}^{(0) \dagger}\\
 &=&
  \begin{pmatrix}
  E_1(\textbf{p}) - \mu & x (\textbf{p})\\
  \bar{x}(\textbf{p}) & -[E_1(\textbf{p}) - \mu ]
 \end{pmatrix}-
 \frac{1}{(E_2(\textbf{p})-\mu)^2 +|y(\textbf{p})|^2}
  \begin{pmatrix}
  0 & u(\textbf{p})\\
  v(\textbf{p}) & 0
 \end{pmatrix}
  \begin{pmatrix}
  E_2(\textbf{p}) - \mu & y(\textbf{p})\\
  \bar{y}(\textbf{p}) & -[E_2(\textbf{p}) - \mu ]
 \end{pmatrix}
 \begin{pmatrix}
  0 & \bar{v}(\textbf{p})\\
  \bar{u}(\textbf{p}) & 0
 \end{pmatrix}\nonumber \\
 &=&
  \begin{pmatrix}
  E_1(\textbf{p}) - \mu & x (\textbf{p})\\
  \bar{x}(\textbf{p}) & -[E_1(\textbf{p}) - \mu ]
 \end{pmatrix}-
 \frac{1}{(E_2(\textbf{p})-\mu)^2 +|y(\textbf{p})|^2}
 \begin{pmatrix}
  -(E_2(\textbf{p})-\mu)^2 |u(\textbf{p})|^2 & u(\textbf{p}) \bar{v}(\textbf{p}) \bar{y}(\textbf{p})\\
  \bar{u}(\textbf{p}) v(\textbf{p}) y(\textbf{p}) & (E_2(\textbf{p})-\mu)^2 |v(\textbf{p})|^2
 \end{pmatrix}\\
 &=&
 \sum_{\mu=0}^3 f_{\mu} (\textbf{p}) \sigma_{\mu}
\end{eqnarray}
with
\begin{eqnarray}
 f_1 (\textbf{p}) -i f_2  (\textbf{p})
 &=& x (\textbf{p})- \frac{u(\textbf{p}) \bar{v}(\textbf{p}) \bar{y}(\textbf{p})}{(E_2(\textbf{p})-\mu)^2 +|y(\textbf{p})|^2} \:, \\
 f_0 (\textbf{p})
 &=&
 \frac{1}{2} \frac{(E_2(\textbf{p})-\mu) (|u(\textbf{p})|^2-|v(\textbf{p})|^2)}{(E_2(\textbf{p})-\mu)^2 +|y(\textbf{p})|^2} \:, \\
 f_3 (\textbf{p})
 &=&
 E_1(\textbf{p})-\mu +\frac{1}{2} \frac{(E_2(\textbf{p})-\mu) (|u(\textbf{p})|^2+|v(\textbf{p})|^2)}{(E_2(\textbf{p})-\mu)^2 +|y(\textbf{p})|^2}
 \:.
\end{eqnarray}
We see clearly that when the time-reversal symmetry is preserved a  BF surface cannot arise, since in this case $\Gamma=\Gamma^\dagger$ which yields $u(\textbf{p})=v(\textbf{p})$, and consequently $f_0 (\textbf{p}) \equiv 0$. The leading order expression of the effective Hamiltonian is found by neglecting the coupling between the heavy particle and hole states, which corresponds to setting $y(\textbf{p}) \rightarrow 0$. This step yields  Eqs.~(17)-(19) in the paper. Since all $x,y,u,v \sim \mathcal{O}(\Delta)$, the leading correction to $f_{0,3}$ is evidently  $\mathcal{O}(\Delta^2)$ whereas the leading correction to $f_{1,2}$ is $\mathcal{O}(\Delta^3)$.

Had we allowed for additional heavy states, the above difference in the orders of corrections to diagonal and off-diagonal terms would apply to all four $2\times 2$ blocks in the $4\times4$ BdG Hamiltonian for the light and the lightest heavy state, after other heavy states are integrated out.
This follows from the (broken) $U(1)$  symmetry in the superconducting phase: if the order parameters $\Delta_a$ change their phases as $\Delta_a \to e^{i \varphi} \Delta_a$, the functions $f_{\mu}$ must transform as:
\begin{eqnarray}
 f_1 - i f_2
 &\to&
 e^{i \phi} \left( f_1 -i  f_2 \right),\\
 f_{0,3} & \to & f_{0,3}
 \:.
\end{eqnarray}
To exhibit this behavior, $f_{1,2}$, or any other off-diagonal term that couples particle and hole states, must be a series of odd powers in the order parameter $\Delta$, and $f_{0,3}$ and other diagonal terms,  a series of even powers of $\Delta$. That is of course what we then find to the leading order in Eqs.~(16)-(19) in  the paper.

The effective Hamiltonian $H_{ef}=\sum_{\mu=0}^3 f_{\mu} (\textbf{p}) \sigma_{\mu}$ has zero energy solutions, when
\begin{equation}
 f_0^2 (\textbf{p}) =f_1^2  (\textbf{p}) +f_2^2   (\textbf{p}) +f_3^2 (\textbf{p})
 \:.
\end{equation}
Hence a necessary condition for the existence of the BF surface is
\begin{equation}
 f_0^2 (\textbf{p}) > f_1^2 (\textbf{p}) +f_2^2  (\textbf{p})
 \:.
\end{equation}
This means that in our two-energy band example the following inequality needs to hold:
\begin{equation}
 \left|\frac{(E_2(\textbf{p})-\mu) (|u(\textbf{p})|^2-|v(\textbf{p})|^2)}{2[(E_2(\textbf{p})-\mu)^2+|y(\textbf{p})|^2]}\right|
 >
 \left| x (\textbf{p})- \frac{u(\textbf{p}) \bar{v}(\textbf{p}) \bar{y}(\textbf{p})}{(E_2(\textbf{p})-\mu)^2 +|y(\textbf{p})|^2} \right|
 \:.
\end{equation}
In the direction where $x(\textbf{p})=\phi^\dagger_1(\textbf{p})\Gamma \phi_1 (\textbf{p}) =0$, it simplifies to
\begin{equation}
 \frac{\Delta}{|E_2(\textbf{p}) - \mu|}
 < \frac{\Delta_c }{|E_2(\textbf{p}) - \mu|}=\frac{1}{B(\textbf{p}) ^{1/2} }
\:,
\end{equation}
where
\begin{equation}
B(\textbf{p})= \bigg[ \frac{2 |u (\textbf{p}) v(\textbf{p}) y (\textbf{p})|}{ \Delta  ( |u(\textbf{p})|^2 - |v(\textbf{p})|^2 )} \bigg] ^2
\:,
\end{equation}
and $B (\textbf{p})\sim \mathcal{O} (\Delta^0)$. The BF surface disappears when the order parameter $\Delta$ reaches the  critical value $\Delta_c$, which is of the order of the difference of the lightest heavy state and the Fermi level, i.e. $\Delta_c \sim |E_2(\textbf{p}) -\mu|$.

The extent $V$ of the BF surface in the ``preferred direction" where  $x(\textbf{p})=0$ is given by $V=2 |f_3 (\textbf{p}) |\sim \Delta^2 / |E_2(\textbf{p})-\mu|$, for $\Delta\ll |E_2(\textbf{p})-\mu|$, and thus proportional to the square of the order parameter divided by the energy of the lightest heavy state, measured from the chemical potential. To determine the maximal extension of the BF surface with $\Delta=\Delta_0$, we study the function $|f_3 (\textbf{p})|=(f_0 (\textbf{p}) ^2 - f_1 (\textbf{p}) ^2 - f_2 (\textbf{p}) ^2 )^{1/2}$, which gives
\begin{equation}
|f_3  (\textbf{p}) |= A(\textbf{p}) \frac{ ( \Delta^2/(E_2 (\textbf{p}) -\mu)) }{1+  C (\textbf{p})  (\Delta^2/(E_2 (\textbf{p}) -\mu)^2 ) }
\bigg[1- B(\textbf{p}) \frac{\Delta^2}{ (E_2 (\textbf{p}) - \mu)^2 }\bigg] ^{1/2},
\end{equation}
where the functions $A(\textbf{p})$ and $C(\textbf{p})$ are also $\sim \mathcal{O} (\Delta^0)$, and read
\begin{equation}
A(\textbf{p}) = \frac{|u(\textbf{p})|^2 - |v(\textbf{p})|^2}{ 2 \Delta^2 },
\end{equation}
\begin{equation}
C(\textbf{p})= \bigg[\frac{y(\textbf{p})}{\Delta} \bigg] ^2.
\end{equation}

The maximum of the last expression for $|f_3 (\textbf{p})|$ is located at $\Delta=\Delta_0$, where
\begin{equation}
\bigg[\frac{\Delta_0}{E_2 (\textbf{p}) -\mu}\bigg]^2 = \frac{[9+ 8 (C(\textbf{p})/B(\textbf{p}))]^{1/2} -3 }{2  C (\textbf{p})}.
\end{equation}
Assuming that the relevant ratio appearing in the above expression,  
\begin{equation}
\frac{ C(\textbf{p}) }{ B(\textbf{p}) } = \bigg(\frac{|u(\textbf{p}) |^2 - |v(\textbf{p}) |^2}{2 |u(\textbf{p}) v(\textbf{p}) |}\bigg)^2,
\end{equation}
is of order of unity, we obtain $(\Delta_0/(E_2 (\textbf{p}) -\mu) )^2 \approx (1/2 B(\textbf{p}))$. In terms of the critical value of the order parameter $\Delta_c$ at which BF surface disappears, the expression for $\Delta_0$ is simply $\Delta_0 ^2 \approx \Delta_c ^2 /2$, as quoted in the text.

\end{widetext}

\end{document}